\documentclass[]{elsart}

\usepackage{epsfig}

\usepackage{amssymb}

\begin{document}

\begin{frontmatter}



\title{Manipulating ionization path in a Stark map: Stringent schemes 
for the selective field ionization in highly excited Rb Rydberg atoms}


\author[ICR]{M. Tada \corauthref{cor}},
\corauth[cor]{Corresponding author.}
\ead{tada@carrack.kuicr.kyoto-u.ac.jp}
\author[ICR]{Y. Kishimoto},
\thanks[now]{
Present address: Research Center for Neutrino Science, Tohoku University, 
Sendai 980-8578, Japan
} 
\author[ICR]{M. Shibata},
\author[ICR]{K. Kominato},
\author[KU]{S. Yamada},
\author[ICR]{T. Haseyama},
\author[ICR]{I. Ogawa \thanksref{now}},
\thanks[now]{
Present address: Department of Physics, Osaka University, Toyonaka, 
Osaka 560-0043, Japan
} 
\author[KU]{H. Funahashi},
\author[KUNE]{K. Yamamoto},
\author[ICR]{S. Matsuki}

\address[ICR]{Nuclear Science Division, Institute for Chemical Research, Kyoto
University, Gokasho, Uji, Kyoto 611-0011, Japan
}
\address[KU]{Department of Physics, Kyoto University, Kyoto 
606-8503, Japan
}
\address[KUNE]{Department of Nuclear Engineering, Kyoto 
University, Kyoto 606-8501, Japan
}

\begin{abstract}
We have developed a quite stringent method in selectivity to ionize the 
low angular-
momentum ($\ell$) states which lie below and above the adjacent 
manifold 
in highly excited Rb Rydberg atoms.  The method fully exploits 
the pulsed field-ionization characteristics of the manifold states in 
high slew-rate regime: Specifically the low $\ell$ state 
below (above) the adjacent manifold is firstly 
transferred to the lowest (highest) state in the manifold via 
the adiabatic 
transition at the first avoided crossing in low slew-rate regime, and then 
the atoms are driven 
to a high electric field for ionization in high slew-rate regime.  
These extreme states of the manifold 
are ionized at quite different fields due to the tunneling  
process, resulting in thus the stringent selectivity.                  
Two manipulation schemes to realize this method  actually are demonstrated 
here experimentally.      
\end{abstract}


\end{frontmatter}


One of the attractive characteristics of the  Rydberg atoms is that 
they  
couple strongly to the external electromagnetic 
field\cite{Stebbings,Gallagher}.  From this property,  
it has been proposed that the atoms are utilized for a single photon 
detector 
in the microwave range\cite{Kleppner,Matsuki} 
besides a number of other applications in fundamental physics 
research  such as the cavity quantum 
electrodynamics~\cite{QED}. In fact the electric 
dipole transition matrix 
elements between the states with principal quantum number $n$ and 
$n \pm 1$ are larger with 
increasing $n$ (proportional to $ n^{2} a_{0}$, where $a_0$ is the 
Bohr radius) 
and yet the lifetime of these states is long enough (proportional to 
$n^{3}$ for the low $\ell$ ($ \ell \ll n$) states) 
to be utilized for this purpose. In 
actual schemes along this line, the Rydberg atoms, excited from the 
initially prepared lower state to a  
upper state by 
absorbing microwave photons, are selectively ionized by means of the field 
ionization.

In order to utilize the Rydberg atoms as a microwave single-photon  
detector, 
it is therefore essential to field-ionize the lower and the upper states 
selectively.  In the Rydberg states with  
$n$ less than 50, where the 
corresponding photons are in the millimeter to sub-millimeter wave region, 
this 
selectivity is mostly achieved by fully exploiting the adiabatic transitions in the 
time evolution of atoms in the Stark map\cite{Stebbings,Gallagher}.  
However this 
ionization scheme has increasing difficulty to achieve good 
selectivity for the higher excited Rydberg states because the difference 
in the ionization field for these states via the adiabatic 
transitions is smaller with increasing $n$. Moreover the non-adiabatic 
process becomes dominant for the higher excited Rydberg states, thus 
the efficiency for the selection is correspondingly reduced with this  
conventional scheme.   

In this Letter we present a new stringent method to selectively ionize 
the highly excited Rydberg atoms with $n$ larger than 90; 
this new method is realized by choosing, as the lower and the upper 
states, two low angular-momentum ($\ell$) states which lie below and 
above the adjacent manifold.  The ionization paths 
are then manipulated for the chosen two states  in the following way.  
Firstly the low $l$ state below (above) the adjacent manifold is  
transferred to the lowest (highest)  state in the manifold through the 
first 
avoided crossing  with a slowly varying (low slew-rate) 
pulsed electric field. Then a pulsed field is applied to drive the 
manifold states into the ionization  
along the non-adiabatic paths in the Stark structure in high slew rate 
regime.  

\begin{figure}
\epsfig{file=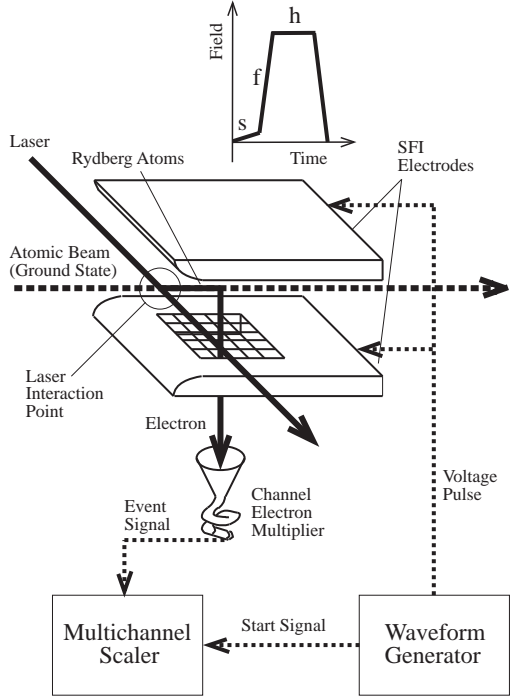,width=7cm}
\caption{Schematic diagram  of the present experimental setup.  
An example of the applied pulse profile is shown 
at the top, in which $s$, $f$ and $h$ denote the components of slow, 
fast and holding time of the pulse, respectively.    
}
\label{fig:setup}
\end{figure}

The stringent selectivity in the above ionization scheme now results from 
the characteristics of the field ionization 
process for the manifold states as revealed in our previous 
work\cite{Kishimoto2002}: In fact the tunneling process in the field 
ionization is increasingly the dominant process in the highly excited 
Rydberg atoms with increasing $n > 90$ \cite{Kishimoto2002}.  The   
ionization 
field value with this process depends strongly on the energy  position 
in the split 
manifold states as well known in Hydrogen atom~\cite{Stebbings,Gallagher}, 
that is, the ionization 
field value is higher in the higher 
energy (bluer)  state than in the lower energy (redder)  
state.  The ionization field values 
for the lowest and the highest states of $n$=120 manifold in Rb, for 
example, are  
4.5 and 10.5 V/cm, respectively, quite large difference compared to 
the case of the adiabatic transition scheme mentioned earlier.      
The effectiveness and universality of the present scheme are  
demonstrated here in the highly excited 
Rydberg states of Rb.\\    
      

Schematic diagram of the present experimental setup 
is shown in Fig.\ \ref{fig:setup}.
A thermal atomic beam of Rb was introduced to the center of the  
field ionization electrodes, where the $^{85}$Rb atoms in the ground 
state $5s_{1/2}$ were excited to the $ns$ and $np$ states  
through the second excited $5p_{3/2}$ state 
by the two-step laser excitation.
A diode laser (780\,nm) 
and a dye laser (479\,nm) with a dye of coumarin 102 pumped by a 
Kr ion laser 
were used for the first and second step excitations, respectively. 
The polarizations of the laser lights adopted in the present experiment 
are 
parallel and perpendicular to the electric field applied for the first and 
the second lasers, respectively\footnote{
No appreciable change in the 
experimental results was observed, however, with  other 
configurations such as both parallel and perpendicular cases. 
}.

The electrodes consist of two parallel plates 
of 52\,mm length and 40\,mm width,
the distance between them being 24\,mm.
A pulsed electric field was applied to the electrodes for the 
field ionization in the 
present experiment. For this purpose a sequence of voltage pulse was 
generated by a waveform generator AWG420 (Sony Tektronix)
and amplified by fast amplifiers before applying to the electrodes.
The pulse profile adopted here has mainly three components, slow 
($s$),  fast ($f$) and holding ($h$) parts as shown in 
Fig.~\ref{fig:setup}; the slow component $s$ is used for 
transferring 
the low $\ell$ states into the manifold states, while the fast 
component $f$ is to ionize the manifold states in high slew rate 
regime as will be explained in detail later.  The component $h$ is 
used for an advanced pulse profile to improve and enhance the selectivity 
more.   The  pulse profile adopted for each measurement will be shown later 
together with the experimental data \footnote{
Repetition rate of the applied pulse was typically 5 kHz.  The rate does not 
give any effect on the experimental results and conclusion presented 
here, but is related 
to the detection efficiency of the Rydberg atoms since we used cw 
atomic beam in the present experiment.  The detection efficiency 
depends on various experimental factors, maximum being about 80 \%.        
}.

The electrons liberated by the field ionization were
guided to a channel electron multiplier
through two fine-mesh grids  placed in one of the electrode plates.
Ionization signals of electrons from the channel electron multiplier
were amplified by a preamplifier and a main amplifier
and then fed to a multichannel scaler P7886 (FAST ComTek)
after the pulse height discrimination.
The ionization events were counted as a function of the elapsed time
from the starting time of the ramp field with the dwell time of 500 ps.
From the correspondence of the applied electric field and the time 
bin, the observed timing spectra were converted into the field 
ionization spectra as a function of the applied electric field. 

\begin{figure}
\epsfig{file=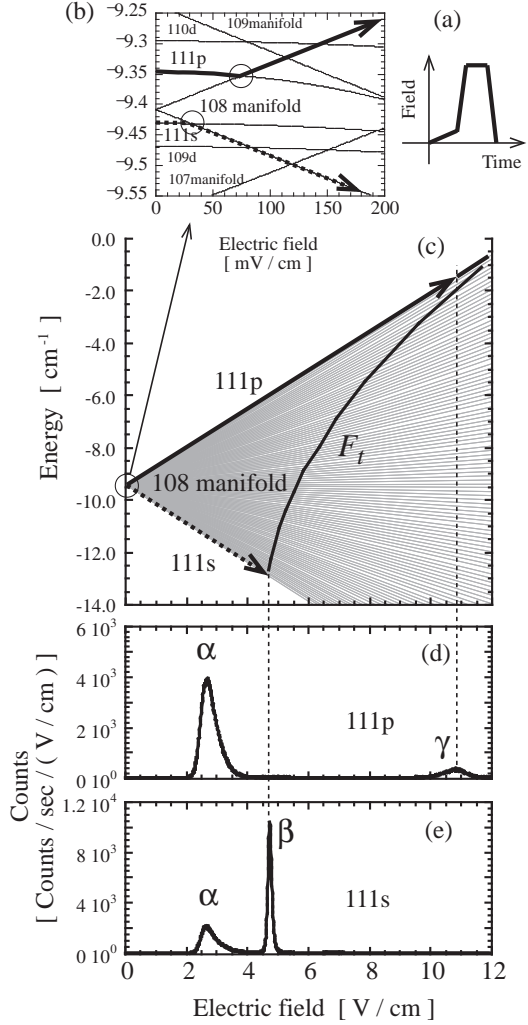,width=7cm}
\caption{Stark energy diagrams near the $n$ = 108 manifold and 
field ionization spectra of the 111$p$ and 111$s$ states in the {\it forward 
driving} scheme.  See the pulse profile (a) here and also the text 
for the meaning of the {\it forward driving}. (a) Schematic 
pulse profile applied. Here 
the slow component is exaggerated to see easily.
(b) and (c) Stark energy 
diagrams near the n=108 manifold; (b) Enlarged 
Stark map at the low electric field, where 
only the lowest and the highest energy states in the manifold are drawn 
for clarity. (c) Stark energy  
diagram covering over the region from zero to the ionization field. 
Solid line ($F_t$) drawn in the split Stark energy levels of 108 manifold is 
the ionization field expected from the tunneling process. Manipulated  
path trajectories in the Stark energy structure  
for the ionization of 111$p$ and 111$s$ states are also shown both in 
(b)  and (c) with thick solid and thick dotted lines, respectively. 
(d) and (e) Field ionization spectra of $p$ and $s$ states in the 
{\it forward driving} scheme; 
(c) 111$p$ state and (d) 111$s$ state.}
\label{fig:pos_drive}
\end{figure}

As the first step for the selective field ionization, two schemes of   
{\it forward} and {\it backward} drivings  for 
the manipulation of the ionization path were examined experimentally;  

1) {\it forward driving} scheme: Typical pulse profile applied for 
this scheme is 
shown in Fig.~\ref{fig:pos_drive} together with the Stark structure 
around $n$ = 108 manifold for $|m_{j}|=1/2$.  The Stark energy 
structure was calculated with the matrix  
diagonalization method\cite{Kishimoto_t}.  The relevant 
energy levels here are the 
111$s$ (lower) and 111$p$ (upper) states which are below 
and above the adjacent 108 manifold, respectively,  as shown in 
the top of the figure.  In order to transfer the initially excited 
$s$ and $p$ states into 
the manifold states, a linearly ramped field is first applied to
the atoms up to $\sim$ 80 mV/cm. Here the 
slew rate of the applied field is slow enough (less than 1 
mV/(cm$\cdot \mu$s))  
to cause the $s$ ($p$) state transfer adiabatically to the lowest 
(highest) manifold state\footnote{
Detailed slew rate dependences of the transition probabilities for the 
adiabatic and non-adiabatic processes at the first avoided crossings 
 of the $s$ (and  also the $p$) state and the adjacent 
manifold states were measured in a separate experiment and compared 
with theoretical predictions taking into account the multilevel-crossing 
effect.  Detailed results of these studies will be presented elsewhere.   
}. 
Then the atoms are driven for 
ionization to high electric field with the same direction to the 
preceding slow component, here in this case up to 14 V/cm,  
with a slew rate of 13.4 V/(cm$\cdot \mu$s).  The observed ionization 
spectra resulting from the applied pulse field are shown also in 
Fig.~\ref{fig:pos_drive}. As presented in detail in our previous 
investigation\cite{Kishimoto2002}, 
 the high field  peaks in the spectra (peaks $\gamma$ and $\beta$ in 
 Figs.~\ref{fig:pos_drive}$d$ and \ref{fig:pos_drive}$e$, respectively) 
 come from  the tunneling process and the values of the 
ionization field corresponding to these peaks 
are quite different. 
We call this scheme {\it forward driving} scheme. 

\begin{figure}
\epsfig{file=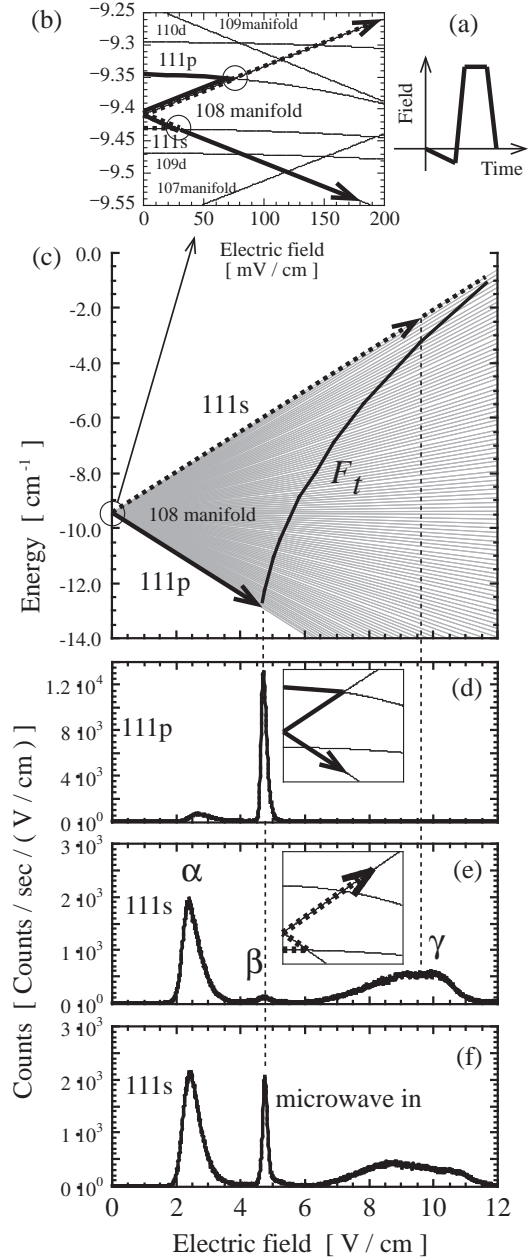,width=7cm}
\caption{Stark energy diagrams near the $n$ = 108 manifold and 
field ionization spectra of the 111$p$ and 111$s$ states in the {\it 
backward driving} scheme.  The applied pulse profile 
is shown at the top right (a). Here 
the slow component is exaggerated to see easily. 
See the pulse profile here and also 
the text for the meaning of the {\it backward driving}.   
(b) and (c) The same Stark energy diagrams near the n=108 
manifold as in Fig.~\ref{fig:pos_drive} except for the manipulated 
path trajectories in the ionization of 111$p$ and 111$s$ states.   
(d) - (f) Field ionization spectra of $p$ and $s$ states in the 
{\it backward driving} scheme. Manipulated  path trajectories at the 
low field region in the 
ionization process are schematically shown in the inset of  
(d) and (e); 
(d) 111$p$ state, (e) 111$s$ state, (f) 111$s$ state initially  
excited with microwave source in.}
\label{fig:neg_drive}
\end{figure}

2) {\it backward driving} scheme: In Fig.~\ref{fig:neg_drive}$a$ shown is 
the pulse profile applied for this scheme, where the initially applied slow 
component and the following fast pulse component of the electric 
field have opposite polarity.  The reversed electric field causes the 
atoms driven to zero field once after the transition to the manifold 
states and then driven to higher electric 
field for ionization.  This reversed ($backward$) driving of the applied 
electric 
field makes the ionization path completely different from the case of 
the $forward$  
driving: In fact the peaks due to the 
tunneling process for the 
$p$ and $s$ states in Figs.~\ref{fig:neg_drive}$d$ and \ref{fig:neg_drive}$e$ 
have reversed 
positions to those shown in Figs.~\ref{fig:pos_drive}$d$ and 
\ref{fig:pos_drive}$e$ in their 
ionization fields.  This kind of reversed driving in 
the ionization process has been previously studied by 
Harmin\cite{Harmin} theoretically and also by several 
groups \cite{ref_rev} experimentally but only at lower $n$ region. 
More detailed characteristics of this {\it backward driving} scheme 
revealed systematically in the present experiment will be presented 
elsewhere.    
%
%

The above two driving schemes have different features as revealed in the 
observed spectra: 1) The fraction of the tunneling-process component 
to the total ionization signals for the 111$p$ state is larger in the 
case of {\it backward  driving}, 2) the width of the  
component for the 111$p$ state in the {\it backward driving} is 
narrower than in the {\it forward driving}.  These features 
indicate that the {\it backward driving} scheme is superior for the 
microwave single-photon detector from the view point of the selectivity 
and the detection efficiency, although the peak position of the 
tunneling-process component for the 111$p$ state is 
in between the two peaks for the 111$s$ state.

\begin{figure}
\epsfig{file=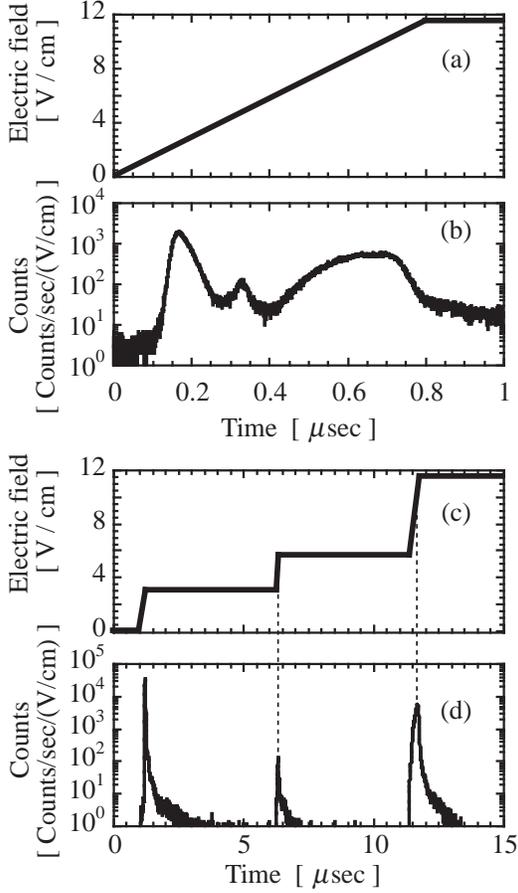,width=7cm}
\caption{
Pulse profile and the corresponding field ionization spectrum of  
the 111$s$ state obtained with     
the advanced {\it three-step} field ionization scheme together with 
those for the standard {\it one-step} scheme presented in 
Fig.~\ref{fig:neg_drive}.  Note here that the slow component of the 
applied pulse in this {\it backward driving} scheme is omitted  
both in the profile drawing of (a) and (c) for simplicity.  
(a) and (b) {\it One-step} pulse profile adopted in the previous 
standard {\it backward 
driving} scheme as presented in Fig.~\ref{fig:neg_drive} and 
corresponding timing spectrum in logarithmic scale.  (c) and (d) Advanced 
{\it three-step} pulse profile in the {\it backward driving} scheme and 
corresponding timing spectrum observed.        
}
\label{fig:3step}
\end{figure}

It is also noted here that a small peak (denoted by $\beta$) 
observed in between the two 
prominent peaks ($\alpha$ and $\gamma$) in 
Fig.~\ref{fig:neg_drive}$e$ in the ionization spectrum of 
111$s$ state is due to the transition induced by the 
blackbody radiations from the initially prepared 111$s$ state to the 
neighboring $p$ states:
In fact  this peak position corresponds 
exactly to that observed in the excitation of the 111$p$ state as seen in 
Fig.~\ref{fig:neg_drive}$d$.  The observed ratio (2\%) of this peak 
component to that of the initially excited 111$s$ component   
 is also in rough agreement with the value (3\%) 
estimated from the relevant transition rates under the actual experimental 
conditions.  
The origin of this peak was further confirmed 
by introducing a microwave source into the present 
experimental system, where the source frequency  is tuned to the 
transition frequency from the 111$s_{1/2}$ to the 111$p_{3/2}$ states; 
in Fig.~\ref{fig:neg_drive}$f$ shown is the enhanced peak due to the 
effect of this microwave source  
\footnote{
Similar transition component would be also expected in the counter 
case of {\it forward driving} scheme.  This component is, however, 
too small to be seen clearly in the spectrum shown in 
Fig.~\ref{fig:pos_drive}$e$ because of its small relative ratio to the 
lower   
peak from the autoionization-like process and also of its broad 
distribution.      
}.
%
%
Manipulating the ionization path as described above was thus found to 
be quite effective to selectively ionize the low $\ell$ states below and 
above the adjacent manifold.  However the lower peak ($\alpha$) due to the 
autoionization-like process is rather broad and has a tail component which 
extends to the region of the peak ($\beta$) due to the blackbody induced 
transitions as noted above.  The extension of this tail part to 
the transition component is more clearly seen 
in Fig.~\ref{fig:3step}$b$ where a raw timing 
spectrum is shown in logarithmic scale.  

In order to improve the selectivity further more,  the applied pulse 
profile was modified as shown in Fig.~\ref{fig:3step}$c$; 
essential 
modification here is that the pulse profile at the 
ionization stage is divided into three 
parts, in each of which the applied 
electric field is held for some period  
before proceeding to the next step of ionization.  Each part of the 
pulse profile is used to induce the field ionization corresponding to 
each peak observed in the previous scheme as seen in 
Fig.~\ref{fig:3step}$b$.    
However this 
division of the pulse profile into three parts is not just for 
dividing the ionization spectra into three in which each of the three 
peaks is just included. Instead   
this pulse profile actually enhances the separation between the 
low field 
ionization (portion $\alpha$ in Fig.~\ref{fig:neg_drive}$e$) of 
the autoionization-like process and the 
following ionization related to the $s$ - $p$ transition (portion 
$\beta$ in Fig.~\ref{fig:neg_drive}$e$) as clarified actually in 
Fig.~\ref{fig:3step}$d$. This is 
because the ionization at the portion $\alpha$ has a finite lifetime, 
which depends  
strongly on the applied electric field \cite{Kishimoto_t}, 
and even at the 
lower electric field than its peak field, the atoms relevant to this   
ionization process are ionized  
within the holding time (5 $\mu$s in the 
present case) longer than their lifetimes. 
Since the ionization field value to be held for this part can be chosen 
in such a way that at this lower field 
the atoms in the upper $p$  state are not ionized at all, the extended 
tail of the lower peak $\alpha$ which entered into the peak region of the 
portion $\beta$ in the previous {\it one-step backward driving} scheme 
is almost completely removed out in this new {\it three-step} scheme.  

The background underneath the transition peak 
was thus reduced by more than two orders of magnitude compared to the 
{\it one-step} scheme described earlier.  
As a result the effective  
noise temperature of this detector system, intrinsically determined 
by the background contributions due to 
the tail contaminations from the initially prepared lower state,  
is well below 10 mK with this scheme.  Thus the present method of 
field ionization detection satisfies the most important  
requirement of low noise characteristics for a sensitive microwave 
single-photon detector.          
     
In conclusion, we have established a quite stringent method to 
selectively ionize the low $\ell$ states which lie below and above the 
adjacent manifold by properly manipulating the ionization path 
through the pulsed electric field.  With more sophisticated pulse 
profile of three steps further, which has the holding time in between 
the steeply 
rising field for ionization, the selectivity is much more enhanced due to the 
characteristics of the ionization mechanism in these highly excited  
Rydberg atoms.  Two actual schemes were proposed and experimentally 
demonstrated to be practical and effective  for 
this 
purpose.  It was shown in particular that the {\it backward driving} 
scheme is more efficient for the Rydberg atoms to be utilized as  
the microwave single photon detector.  The present method  thus opens a new 
way to detect single microwave photons individually over the region of 
10 cm wavelength with higher $n$ Rydberg atoms\cite{Matsuki,Tada}.

This research was partly supported  by a Grant-in-aid for Specially Promoted 
Research by the Ministry of Education, Science, 
Sports, and Culture, Japan (No. 09102010).  M.T., S.Y., and T.H. thank the 
financial support of JSPS, Japan  under the Research Fellowships for the 
Young Scientists. 

\end{document}